\documentclass[aps,twocolumn,prb]{revtex4}
\usepackage{amssymb}

\usepackage{amsmath}
\usepackage[mathcal]{eucal}
\usepackage{graphicx}
\usepackage{dcolumn}
\usepackage{amsfonts}
\usepackage{bm}
\usepackage{array}
\usepackage{natbib}

\newcommand{\be}{\begin{equation}}
\newcommand{\ee}{\end{equation}}
\newcommand{\bea}{\begin{eqnarray}}
\newcommand{\eea}{\end{eqnarray}}

\renewcommand{\vec}[1]{{\bm #1}}

\renewcommand{\Im}{\mathrm{Im}}

\bibpunct{[}{]}{,\!}{n}{,}{,} 
\begin{document}
\title{Spin polarization oscillations without spin precession: spin-orbit entangled resonances in quasi-one-dimensional spin transport}
\author{D. H. Berman}
\affiliation{Department of Physics and Astronomy, University of Iowa, Iowa City, Iowa 52242, USA}
\author{M. Khodas}
\affiliation{Department of Physics and Astronomy, University of Iowa, Iowa City, Iowa 52242, USA}
\author{M. E. Flatt\'e}
\affiliation{Department of Physics and Astronomy, University of Iowa, Iowa City, Iowa 52242, USA}
\begin{abstract}
Resonant behavior involving spin-orbit entangled states occurs for spin transport along a narrow channel defined in a two-dimensional electron gas, including an apparent rapid relaxation of the spin polarization for special values of the channel width and applied magnetic field (so-called ballistic spin resonance). 
A fully quantum mechanical theory for transport through multiple subbands of the one-dimensional system provides the dependence of the spin transport on the applied magnetic field and channel width,
including a resonant depolarization of spins when the Zeeman energy matches the subband energy splittings and a spin texture transverse to the magnetic field. The resonance phenomenon is robust to disorder. 
\end{abstract}

\date{\today }
\maketitle
\section{Introduction}\label{Introduction}
Spin transport in small nonmagnetic structures is an essential element of semiconductor spintronics~\cite{Wolf2001,Awschalom2007}, both for controlling the properties of spin-orbit (SO) fields originating from a material's electronic structure~\cite{Meier1984} and for scaling down spin-dependent devices to the nanoscale~\cite{Hall2006,Heedt2012}. 
SO fields in nonmagnetic semiconductors cause a carrier's spin to precess around an axis with a rate that depends strongly on  that carrier's momentum; an effect which produces coherent spin precession even during diffusive transport~\cite{Kato2004c,Kato2005b,Crooker2005a}.  
The spin of each carrier also dephases relative to other carriers with different momenta; as the carriers scatter from one momentum state to another this dephasing results in decoherence through randomization of the carrier momentum~\cite{Dyakonov1972,Meier1984,Lau2001}. Confinement of the carrier wave function to a smaller region, or application of a magnetic field~\cite{Dyakonov1986,Ivchenko1997}, can quench the carrier momentum along one or more axes, and thus reduce the random character of the SO field. 
Thus confinement of carriers in a semiconductor into a one-dimensional channel was predicted to suppress spin decoherence from such SO fields~\cite{Bournel1998,Malshukov2000,Kiselev2000,Pareek2002}, with the spin decoherence rate approaching zero for a very narrow channel~\cite{Schwab2006}.
Surprisingly, suppression of spin decoherence was measured in channels that were considerably wider than expected~\cite{Holleitner2006}, in which many subbands of the channel were occupied. Furthermore, in specially-designed channels where the spin precession time approximated the transverse transit time of a carrier, an {\it apparent} reduction in the spin relaxation length was observed~\cite{Frolov2009a}, so-called ballistic spin resonance (BSR), suggesting that confinement and applied magnetic fields may instead enhance spin decoherence.

Here these seemingly contradictory statements are reconciled using a quantum-mechanical microscopic theory of spin transport in a channel within which multiple subbands are occupied. The feature identified as BSR in Ref.~\onlinecite{Frolov2009a} originates from coupling, via the  SO interaction, between pairs of subbands that differ in both spin and orbital quantum numbers, yielding coherent precession of a ``pseudospin''. The magnetic-field dependence of the resulting propagating spin polarization is calculated assuming the spins are injected into the channel uniformly across the width of the channel.  The pseudospin precession then yields spatial oscillations of the spin polarization as the distance along the channel from the injection location increases, and a transverse spin texture with no net transverse spin polarization.  
In Ref.~\onlinecite{Frolov2009a} the detector was insensitive to the transverse spin polarization, and was set a fixed distance from the injection point; for such a geometry the BSR phenomenon manifests as an apparent reduction of the spin relaxation length.
We consider the effect of disorder on the propagation and find that the predicted features of spin transport in a narrow channel are robust. Greater spatial resolution in the spin detection process would allow the oscillations to be directly imaged, and could permit them to be used in processing spin information within a small semiconductor device, or guiding it along a controllable pathway within a semiconductor chip.

The fundamental properties of spin transport in the quasi-one-dimensional channel (pseudospin precession and  resonant depolarization of spin) can be derived from a simplified two-subband model. We begin by considering a channel with only two channel subbands,  with dispersion relations shown below in Fig.~\ref{2subband}(a) in the presence of a magnetic field oriented along the $\hat z$ direction. Once the SO interaction is added, subband $1$ with spin down is coupled to subband $2$ with spin up. The result, shown in Fig.~\ref{2subband}(b) are two eigenstates with different energies (upper and lower) which  consist of a mix of up and down spins and the two subband states (orbital motion). When a spin density is injected into the system at the Fermi energy, instead of going into unoccupied spin eigenstates at the Fermi energy, that spin must occupy linear superpositions of the upper and lower mixed eigenstates shown in Fig.~\ref{2subband}(b), which propagate with different wave vectors 
down the channel, as shown.

\begin{figure}[h]
\begin{center}
\includegraphics[width=1.0\columnwidth]{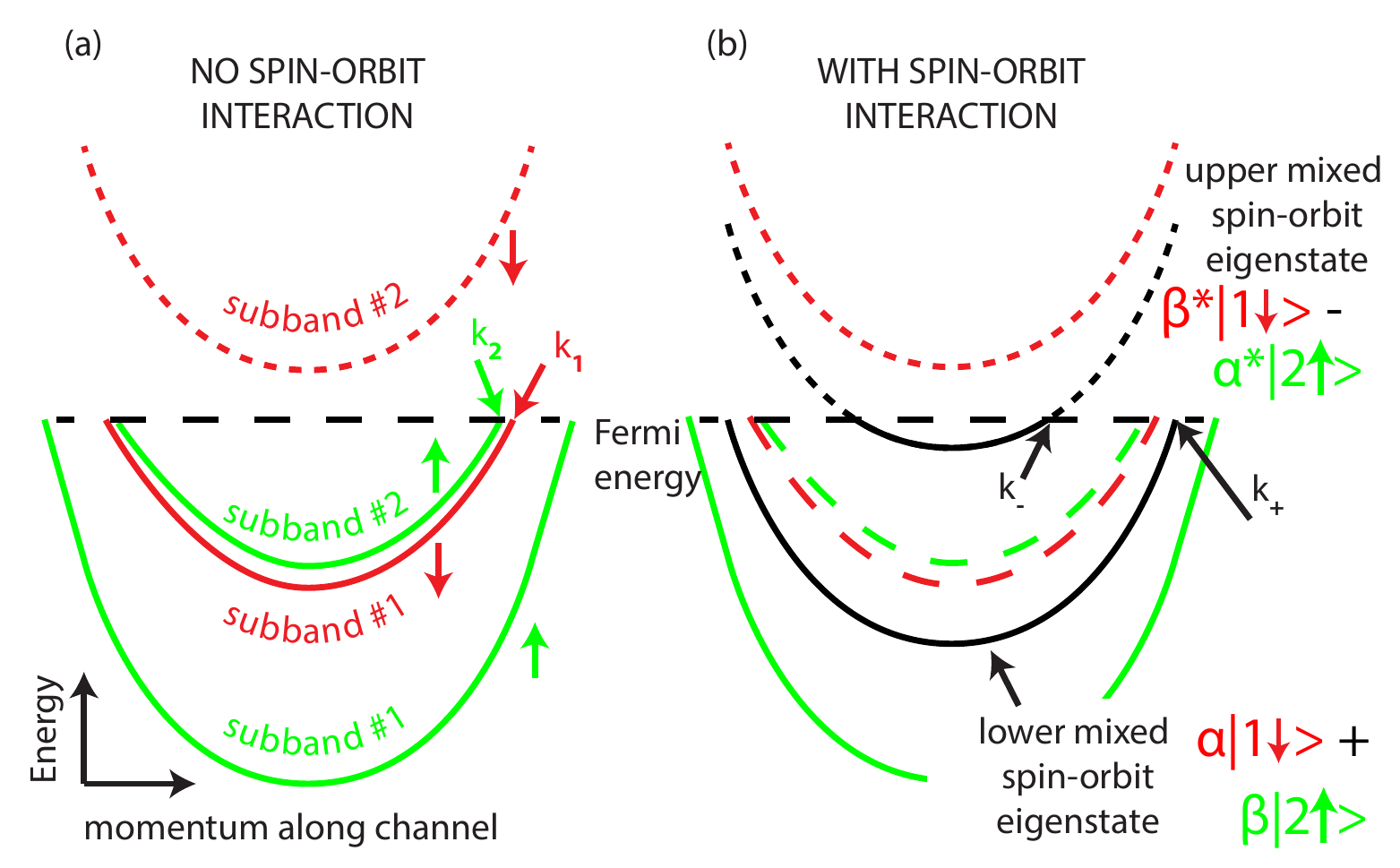}
\caption{ The effect of SO coupling illustrated for two subbands (1 and 2) with different lateral orbital wave functions. 
Without SO the in-plane magnetic field splits each subband into spin-up and spin-down subbands separated by the Zeeman energy  $E_Z$.
At the resonance condition, Eq.~\eqref{Resonant} the spin-down subband of band 1 coincides with the up-spin subband  of band 2.
The SO-coupling-induced subband mixing is strongest exactly at the resonance condition, where $|\alpha| = |\beta| = 1/ \sqrt{2}$.}
\label{2subband}
\end{center}
\end{figure}

\begin{figure}[h]
\begin{center}
\includegraphics[width=1.0\columnwidth]{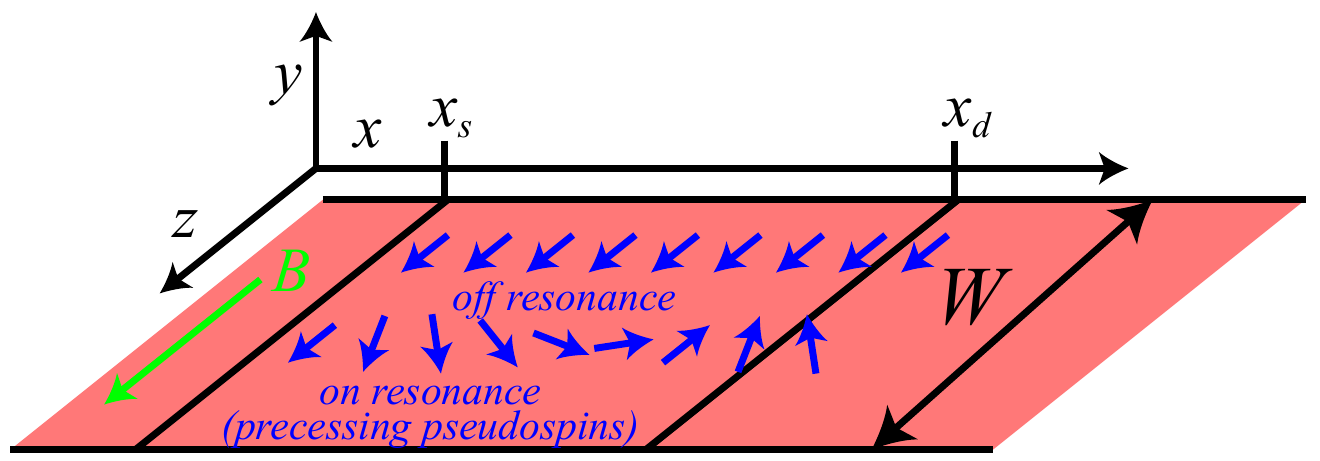}
\caption{ 
Schematic representation of experimental setup of Ref.~\cite{Frolov2009a}.
A quasi-one-dimensional channel of width $W$ is defined by confinement of a two-dimensional electron gas in the $xz$ plane. 
The in-plane magnetic field is applied in the $\hat{z}$ direction.
$x_s$ and $x_d$ denote the location of the source and drain of spin polarization along the channel.
 }
\label{setup}
\end{center}
\end{figure}

Figure~\ref{setup} is a schematic  of Ref.~\onlinecite{Frolov2009a} to help
illustrate the consequences of the spin-orbit mixing of Fig.~\ref{2subband} for quasi-one-dimensional spin propagation. A channel in a two-dimensional electron gas has width $W$ in the $\hat z$ direction and extends along the $\hat x$ direction. The external applied magnetic field is oriented along $\hat z$ so that spins polarized along $\hat z$ are injected and detected by  two quantum point contacts (QPC) located at $x_s$ and $x_d$. 
Without the spin-orbit interaction the two states at the Fermi energy [Fig.~\ref{2subband}(a)] moving towards $+\hat x$ (Fig.~\ref{setup}) are $\exp(ik_1x)\phi_1(z)\eta_\downarrow $ and $\exp(ik_2x)\phi_2(z)\eta_\uparrow$, where $\eta_s$ is a spinor. 
A term in the Hamiltonian such as the Rashba interaction, $\alpha p_z \sigma_x$, will couple these two subbands via the matrix element 
$\left\langle 2,\uparrow |\alpha p_z \sigma_x | 1, \downarrow\right\rangle \ne 0$. The eigenstates resulting from this coupling (denoted by $+$ or $-$) are linear combinations of the nearly degenerate uncoupled states [with wave vectors $k_1$ and $k_2$ indicated in Fig.~\ref{2subband}(a)]. They form spin-orbit-entangled states. As indicated in Fig.~\ref{2subband}(b) these entangled states split in energy at fixed wave vectors and they cross the Fermi energy at different wave vectors, $k_+$ and $k_-$. 


For the geometry of Fig.~\ref{setup} the net (integrated along the transverse ($z$) axis of the channel) fractional density of 
spins polarized in the $z$-direction 
at position $x$ along the channel, arising from $z$-polarized spins   injected  into the channel at $x_s$, is
\begin{align}\label{res:single}
\tilde s_z(x,B) = 
  \cos^2 \theta\! + \!\sin^2 \theta  \cos \left[ (k_+ - k_-) (x - x_s) \right] 
,
\end{align}
where  
$(k_+ - k_-)$ is the difference of the wave vectors parallel to the channel for subbands $1$ and $2$ in the presence of spin-orbit interaction, and $(k_+ - k_-)^{-1}$ provides the length scale of the spin polarization oscillation. The value of $\theta$ determines the system's nearness to resonance, and is quantified in Sec.~\ref{sec:Spin}.
We will refer to the part of $s_z(x,B)$  that does not oscillate with $x$ as the conserved part of the spin density.
The oscillation in $x$ can be understood as resulting from an effective precession. If the energy difference of two modes at $k_1=k_2$ is $(E_+ - E_-)$, then  the precession expresses the beating of the two phases in time:
$$  (E_+ - E_-) t/\hbar = \partial_k E  (k_+ - k_-) t/\hbar = $$
\begin{equation}\label{2} (k_+ - k_-) v t = (k_+ - k_-)(x -x_s) .\end{equation}

The spins behave {\em as if} they encounter an effective magnetic field  ${\mathbf B}_{\mathrm{eff}}$ in the direction ${\hat e} = ( \sin \theta, 0 , \cos \theta) $ and the spins precess about this  field as they progress down the channel, however 
the absence of a net spin precession is clearly indicated by the result that $s_x(x,B) = 0$ everywhere along the channel.    The transverse spin polarization does not vanish, however, but instead has a texture $s_x(x,z,B)\propto \phi_1(z)\phi_2(z)$, determined  by the product of the orbital wave functions $\phi_1(z)$ and $\phi_2(z)$.

\begin{figure}[h]
\begin{center}
\includegraphics[width=1.0\columnwidth]{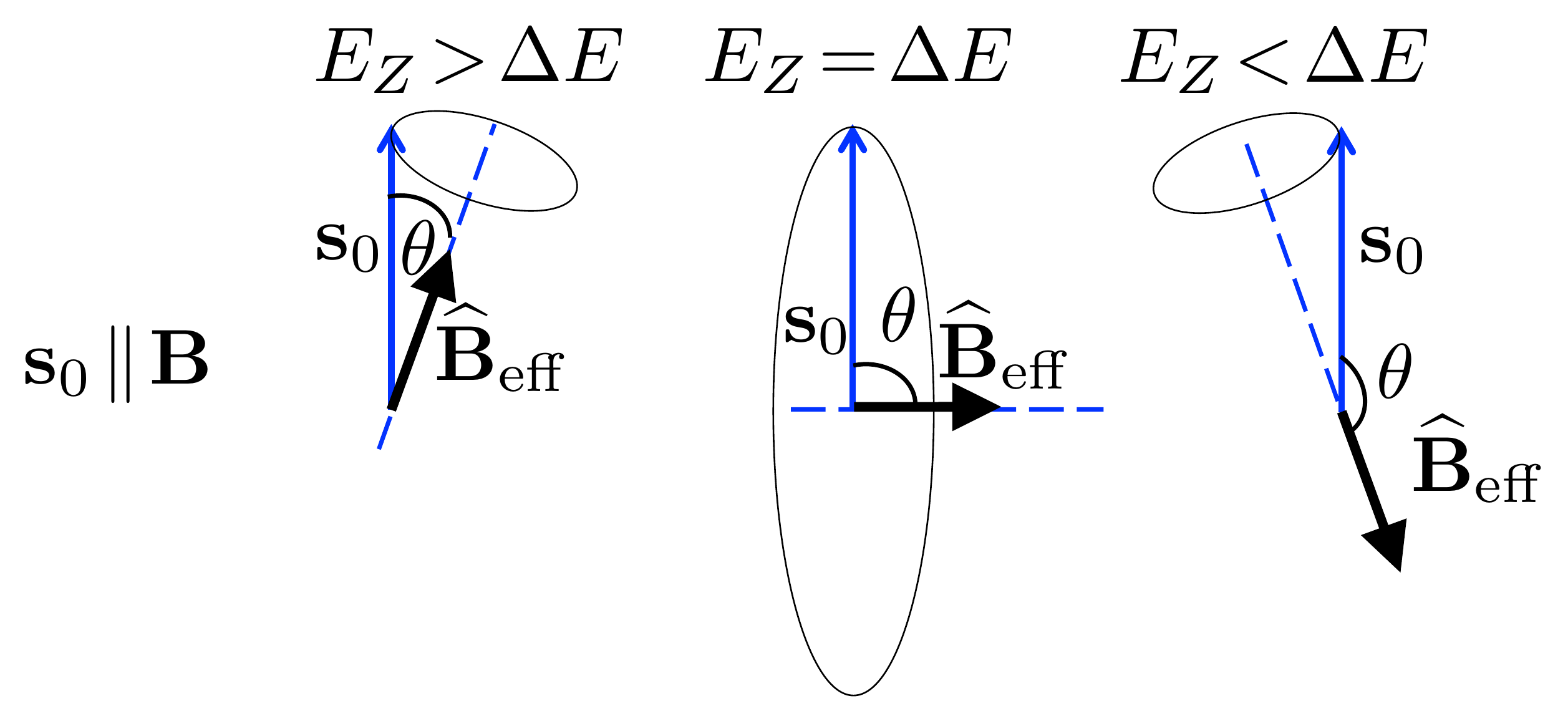}
\caption{ The injected spin polarization, $\vec{s}_0$ is directed along the external in-plane magnetic field $\mathbf{B}\parallel \hat{z}$, see Fig.~\ref{setup}. 
The effective magnetic field, $\mathbf{B}_{\mathrm{eff}}$ determines the spin precession after the injection.
The $z$-component of $\mathbf{B}_{\mathrm{eff}}$ given by Eq.~\eqref{Beff_z} vanishes at the resonance, Eq.~\eqref{Resonant}.
In this case the spin dynamics is a pure precession with a frequency controlled by the SO coupling matrix element, $(g \mu_B/\hbar)  \left| \mathbf{B}_{\mathrm{eff}}\right|  =  (1/\hbar)\left| \left\langle n+1,\uparrow |\alpha p_z \sigma_x | n, \downarrow\right\rangle \right|$.
Off the resonance a finite component of injected spin ($\propto \cos \theta$) is conserved and is represented by an $x$-independent part in Eq.~\eqref{res:single}.
 }
\label{precession}
\end{center}
\end{figure}

The effective field  in the $x$ direction, ${ B}_{\mathrm{eff},x}$,  arises from the SO matrix elements $ \left\langle n+1,\uparrow |\alpha p_z \sigma_x | n, \downarrow\right\rangle$ between non-SO coupled states. The { effective} magnetic field in the
$z$ direction arises from energy splitting between adjacent modes because of both confinement ($\Delta E$)  and  the Zeeman interaction with the actual external field ${ B}_z$,
\begin{equation}\label{Beff_z}
 { B}_{\mathrm{eff},z}  =  { B}_z - \Delta E/( g \mu_B) .
\end{equation} 
Ballistic spin resonance then occurs when  ${ B}_{\mathrm{eff},z} = 0$, corresponding to
\begin{equation} 
 g |\mu_B| B_z \equiv E_Z =  \Delta E, 
 \label{Resonant} 
 \end{equation}  
and the effective field lies entirely in the direction along the channel, so spins in the $z$-direction have no non-precessing part, as schematically indicated in Fig.~\ref{precession}. 
Equation \eqref{Resonant} is  the same condition for resonance as cited in Ref.~\onlinecite{Frolov2009a}.  For square-well confinement only a few  modes near cut-off can be in resonance for a given external field $B$;   for parabolic confinement adjacent modes at $B=0$ are all separated by the one energy $\hbar \omega=\Delta E$,  so that all pairs go into resonance for the same magnetic field.

In the next section we derive a general expression for the spin density, which is exact provided the Rashba and Dresselhaus coefficients are of opposite sign and equal magnitude, and provided there is neither  scattering nor electron-electron interaction; Eq.~\eqref{res:single}  is a special case of this expression. 
In Sec.~\ref{sec:Spin} we compare spin densities when a) the confining potential is parabolic and b) when it simulated by an infinite square well. 
In Sec.~\ref{sec:Effect} we study the effects of  scattering from disorder on the spin-density.  
In Sec.~\ref{sec:Discussion} we further discuss our results and their implications.

\section{Theory of spin polarization transport in a quasi-one-dimensional channel}
\label{sec:Spin}

We now generalize to the case where multiple subbands are occupied, and derive a general result motivated by Eq.~(\ref{res:single}).  Let the rate of the spin injection per unit length be $F(x)$.
Then the stationary polarization is given by
\be\label{Main3}
s_z (x,B) = 
\left.
- i \int \frac{ d q }{ 2 \pi }e^{ i q x }  
\left[ \partial_{\omega} \chi(q,\omega) \chi^{-1}(q) \right] \right|_{\omega \rightarrow i 0} F(q) \, .
\end{equation}
Equation \eqref{Main3} relates the spin density to the Fourier image of the injection rate $F(q)$ via the retarded correlation function
\be\label{retcorr}
\chi(q,\omega) =  
- i \int_0^{\infty}\! d t\! \int\! \! d x
e^{ - i q x + i \omega t }
\langle [\hat{s}_{z}(x,t),\hat{s}_{z}(0,0)] \rangle
\ee
of the spin density operator,
\be\label{spin_op}
\hat{s}_z(x) = \frac{1}{2} \int_{-\infty}^{+\infty} d z 
\left[ \psi^{\dag}_{\uparrow} (x,z) \psi_{\uparrow}(x,z)  - \psi^{\dag}_{\downarrow}  (x,z ) \psi_{\downarrow}  (x,z)
\right]
\ee
where $\psi^{\dag}_{\uparrow,\downarrow}(x,z)$ is the second quantized creation operator and the spin quantization axis $\hat{z}$ coincides with the direction of the in-plane magnetic field which is perpendicular to the channel, as shown in Fig.~\ref{setup}.  Equation~\eqref{Main3} is derived in the Appendix, and its limitations are discussed in Sec.~\ref{sec:Discussion}.

We apply Eq.~\eqref{Main3} to non-interacting electrons in a clean wire with parabolic or square-well confinement, with the 
Hamiltonian
\be\label{H_clean}
H = \frac{p_x^2 + p_z^2 }{2 m } + V_c(z)  - \frac{1}{2} E_Z \sigma_z  + H_{SO}\, .
\ee
Here $V_c(z)$ is the lateral confinement potential,
$E_Z $ is the Zeeman splitting, and $H_{SO} = \alpha_- p_z \sigma_x + \alpha_+ p_x \sigma_z$ is the SO interaction term with  $\alpha_{+,(-)} $ equal to the sum (difference) of the Rashba and Dresselhaus coefficients. As the term with $\alpha_+$ is proportional to $\sigma_z$, it does not couple subbands with opposite spin direction along $\hat z$; its effect is entirely to provide a $k_x$ dependent shift of the Zeeman energy. Therefore to simplify the calculations presented here, we will neglect the effect of this term and set $\alpha_+ = 0$ throughout this paper.

Without SO coupling the eigenstates of the Hamiltonian, Eq.~\eqref{H_clean},
can be denoted by their
wave vector 
along the channel, $k$, their spin $s$ (up or down) in the z-direction across the channel and the natural mode numbers $m$ arising from confinement: 
\begin{equation} 
\langle x,z |k, m, s \rangle  = \frac{e^{i k x}}{\sqrt{L}} \phi_{m}(z) | s \rangle. \end{equation}
The eigenstates with SO coupling can be expanded in terms of these uncoupled states as 
\begin{equation}\label{ball_eig}
|k,n,\gamma \rangle = \sum_{m,s} A^{n,\gamma}_{m,s}( B )  | k, m, s \rangle,
\end{equation}
($\gamma = \pm 1 $)
with corresponding energies given by 
\begin{equation} \epsilon_{k,n,\gamma} = \frac{\hbar^2 k^2}{2 m} + \epsilon_{n,\gamma} .
\end{equation} It is because $\alpha_+ = 0$ that the expansion coefficients $A^{n,\gamma}_{m,s}( B )$ do 
not depend on the momentum $k$ and the energies depend only on $k^2$.
We stress that the eigenstates $|n,k,\gamma \rangle$ in Eq.~\eqref{ball_eig} are  superpositions of states with different subband index $m$ and spin index $s$.
Such mixing is due to SO coupling and is shown in Fig.~\ref{2subband} for the case of two subbands.

The retarded susceptibility  is given in terms of the eigenstates and energies of the coupled Hamiltonian by
\begin{align}
\chi(q,\omega) = \frac{- 1}{2 \pi} \sum_{n,\gamma,n',\gamma'}  
| \langle n,\gamma | \sigma_z | n',\gamma' \rangle |^2    \notag \\
\times   \int d k 
\frac{   f( \epsilon_{k,n,\gamma} ) - f( \epsilon_{k +q ,n',\gamma'} )}
{ \omega+ i 0 + \epsilon_{n,\gamma} -  \epsilon_
{n',\gamma'} - (\frac{\hbar^2}{2 m})
( 2 k q +q^2) } . 
\label{eigens}
\end{align}  
The zero temperature Fermi
function is   $ f(\epsilon) = \Theta ( \mu - \epsilon ) $, where $\Theta$ is
the Heaviside function and $\mu$ is the chemical potential.
The integration over $k$ can be 
performed analytically. The static susceptibility is then
 \begin{align} \chi^s(q)   =   \frac{m}{ \pi \hbar^2 q}   \sum_{n,\gamma,n',\gamma'} 
| \langle n,\gamma | \sigma_z | n',\gamma' \rangle |^2  \notag \\
\times\log \frac{ |q + ( k_{n,\gamma} + k_{n',\gamma'})|}{ |q- ( k_{n,\gamma} + k_{n',\gamma'})|}. \label{chistatic1}
\end{align} 
When $q \rightarrow 0$,
\begin{equation} \chi^s ( q = 0 ) = 
 \frac{2 m}{ \pi \hbar^2 }   \sum_{n,\gamma,n',\gamma'} \frac{
| \langle n,\gamma | \sigma_z | n',\gamma' \rangle |^2  }{ k_{n,\gamma} + k_{n',\gamma'} }\, ,
\end{equation}
where $k_{n,\gamma} = \sqrt{ (2 m/\hbar^2 )(\mu - \epsilon_{n,\gamma})}$ is the wavenumber for propagation along the wire. 
The sum over modes is restricted to those with real $k_{n,\gamma} .$

To compute the spin density we will also need the frequency derivative of the 
susceptibility.  The integrals here can also be computed with the result
\begin{align} \label{Dchi}
 \partial_\omega & \chi( q, \omega + i\delta)|_{\omega=0}  =    
 \frac{i2 m^2 }{ \hbar^3 }\sum_{n,\gamma,n',\gamma'} | \langle n,\gamma | \sigma_z | n',\gamma' \rangle |^2 
   \notag \\
 \times &  \left[  \frac{\delta(  q + k_{n,\gamma} -  k_{n',\gamma'})}{2 k_{n,\gamma} k_{n',\gamma'}}  + 
 \frac{\delta(  q + k_{n,\gamma} +  k_{n',\gamma'})}{ 2k_{n',\gamma'}( k_{n,\gamma} +  k_{n',\gamma'})}  \right.   \notag \\ 
 & + \left. \frac{\delta(  q - k_{n,\gamma} -  k_{n',\gamma'})}{ 2k_{n',\gamma'}( k_{n,\gamma} +  k_{n',\gamma'})}  \right] \, .  
 \end{align}
Finally we  arrive at a fairly simple, yet exact, result for the spin density using
Eqs.~\eqref{Main3}, \eqref{chistatic1} and \eqref{Dchi},
\begin{align}
s(x) & =  s_0  \frac{ m^2}{2 \pi \hbar^3} 
\notag \\
\times &\sum_{n,\gamma,n',\gamma'} 
\frac{| \langle n,\gamma | \sigma_z | n',\gamma' \rangle |^2 \cos[ (k_{n,\gamma} -  k_{n',\gamma'}) (x-x_0) ]}{ \chi^s( k_{n,\gamma}-k_{n',\gamma'})( k_{n,\gamma}k_{n',\gamma'})}. \label{sExact}
\end{align}

For a  source located at $x=x_0$ we  take $F(q) =s_0 \exp(-i q x_0)$.
In practice the matrix elements are only large when $(n,\gamma)$ is ``near" 
$(n',\gamma')$  and the SO splitting is small so that $k_{n,\gamma}-k_{n',\gamma'}$ is small.  Then no major error is incurred if in the static susceptibility we set $k_{n,\gamma}-k_{n',\gamma'} =0 $.  It should be noted too, that in going from Eq.~\eqref{Main3} to Eq.~\eqref{sExact}, we made use of the fact that the static susceptibility diverges (logarithmically) when $q = \pm ( k_{n,\gamma}+k_{n',\gamma'})$ for any pair $ (n,\gamma)$ and  $ (n',\gamma')$, so that the 
corresponding $\delta$ functions in the derivative of $\chi$ do not contribute to 
the spin density. This result for the spin density is exact, depending only on the 
the assumptions that the wire is clean, the particles only interact with the confining potential, spin-orbit field and magnetic field, and $\alpha_+ = 0$.  
Diagonal terms in the double sum over states give a
contribution to the spin density 
independent of the distance $x-x_0$, 
while off-diagonal contributions provide a sum of sinusoidal contributions with spatial beat frequencies $k_{n,\gamma} - k_{n',\gamma'}$.  
Equation \eqref{sExact} becomes  plausible when 
the modal expansion of  the Green's function in the channel is recalled: $$ G \propto  \sum_{n,\gamma} 
| n,\gamma \rangle \langle n, \gamma |  \frac{ e^{i k_{n,\gamma} |x - x_0| }}{ k_{n,\gamma}} . $$

The result for the spin density cited at the outset can now be obtained for parabolic confinement with $V(z) = m \omega^2 z^2/2$  . Then the only non-zero matrix elements are
\begin{align}\label{cos_n}
  | \langle n,\gamma | \sigma_z | n,\gamma \rangle |^2  = \cos^2 \theta_n =  \left(\frac{E_Z -\hbar \omega}{\Delta_n }    \right)^2 
\end{align}
and 
\begin{align}\label{sin_n}
 | \langle n,\gamma | \sigma_z | n,-\gamma \rangle |^2  = \sin^2 \theta_n =  \left(  \frac{\delta_n^{S0}}{\Delta_n}   \right)^2\, ,
 \end{align}
where the SO splitting is given by
\begin{equation}\label{Delta_n^SO}
\delta_n^{SO} = \alpha_- \sqrt{ 2 ( n + 1) m \omega/ \hbar } 
\end{equation}
and the level splitting by
\begin{equation}\label{Delta_n}
\Delta_n = \sqrt{ (\delta_n^{SO})^2 + (E_Z - \hbar \omega)^2 } \, .
\end{equation}
The simplicity of the result for parabolic confinement occurs in part because the momentum operator for the harmonic oscillator has matrix elements only between adjacent non-SO coupled states. 

For parabolic confinement we find then
\begin{align}\label{res:clean}
s_z(x,B) = & \frac{s_0}{2 \sum_{n} v_n^{-1}} \sum_{n} 
 v_n^{-2 } \Bigg\{ 
\cos^2 \theta_n 
\notag \\
&
+ \sin^2 \theta_n  \cos \left[ \frac{2 \pi}{ l_n } (x - x_s) \right] \Bigg\} \, , 
\end{align}
where the sum runs over pairs of propagating modes,  and $2 \pi l_n^{-1} = (k_{n,1} - k_{n,-1})  $ is $2\pi$ times the inverse spin precession length of an electron in  the $n$th modes.
$k_{n 1}$
and $k_{n,-1}$ have been replaced with $( k_{n, 1} + k_{n, -1})/2 = m v_n/ \hbar$ except in the term oscillating with $x-x_s$.
The result for a single mode is Eq.~(\ref{res:single}).
For other confining potentials, such as an infinite square well, eigenstates and energies can be found by diagonalizing a matrix with dimensions only slightly larger than the number of propagating modes.

 \begin{figure}
 \includegraphics[width=1.0\columnwidth]{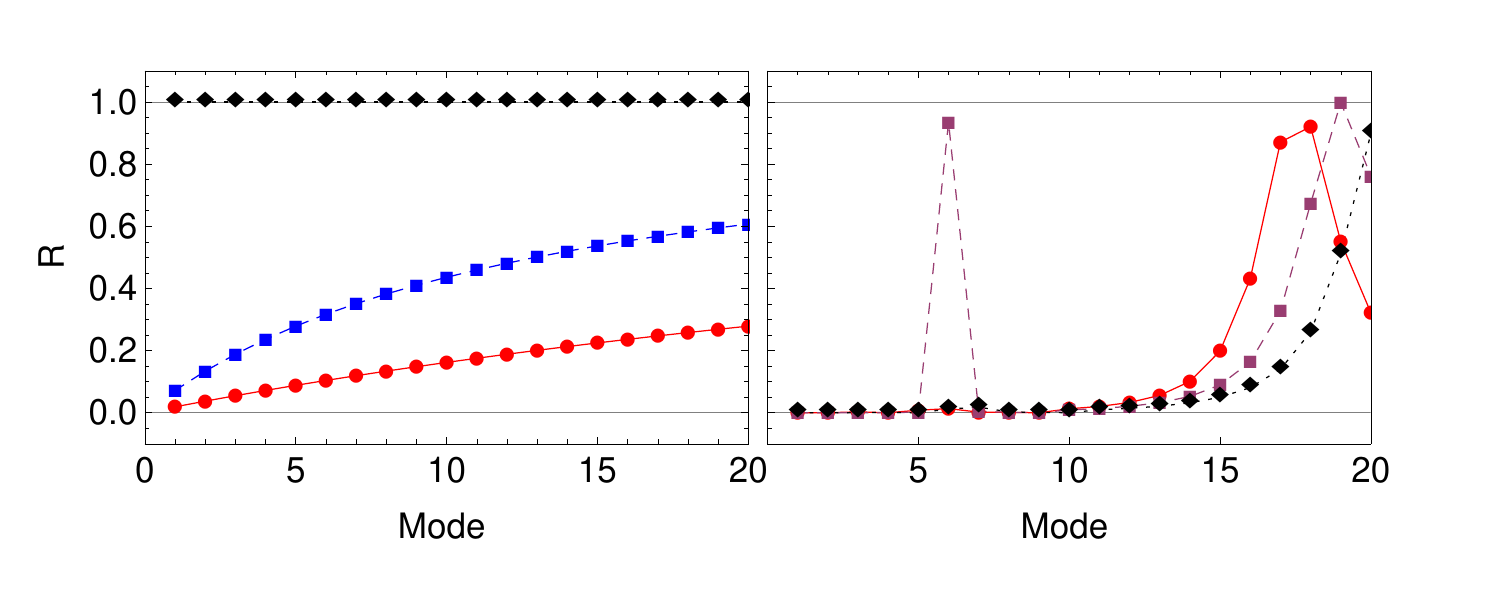}
 \caption{(Color online) Ratio of adjacent energy level splitting due to SO interaction to the total  energy level splitting for pairs of propagating modes, $R =   |\langle n,\gamma | \hat{s}_z  | n,-\gamma \rangle |^2/ (  |\langle  n,\gamma | \hat{s}_z  |n,\gamma \rangle |^2  + |\langle n,\gamma | \hat{s}_z  |n, -\gamma  \rangle |^2 ). $  Parameters are chosen so that $B_{resonant} = 7 $ T and so that there are 20 propagating modes; W = 1.15 $\mu$m.  In the parabolic case $R$ is just $\sin^2 \theta_n.$ The left panel shows
  R for parabolic confinement; right panel shows R for square well confinement.  B = 3 T is  (solid, red) , B= 6 T (dashed, blue) and B = 7 T (dotted, black)  . }
  \label{fig:matrix_elements}
 \end{figure}

 \begin{figure}
 \includegraphics[width=1.2\columnwidth]{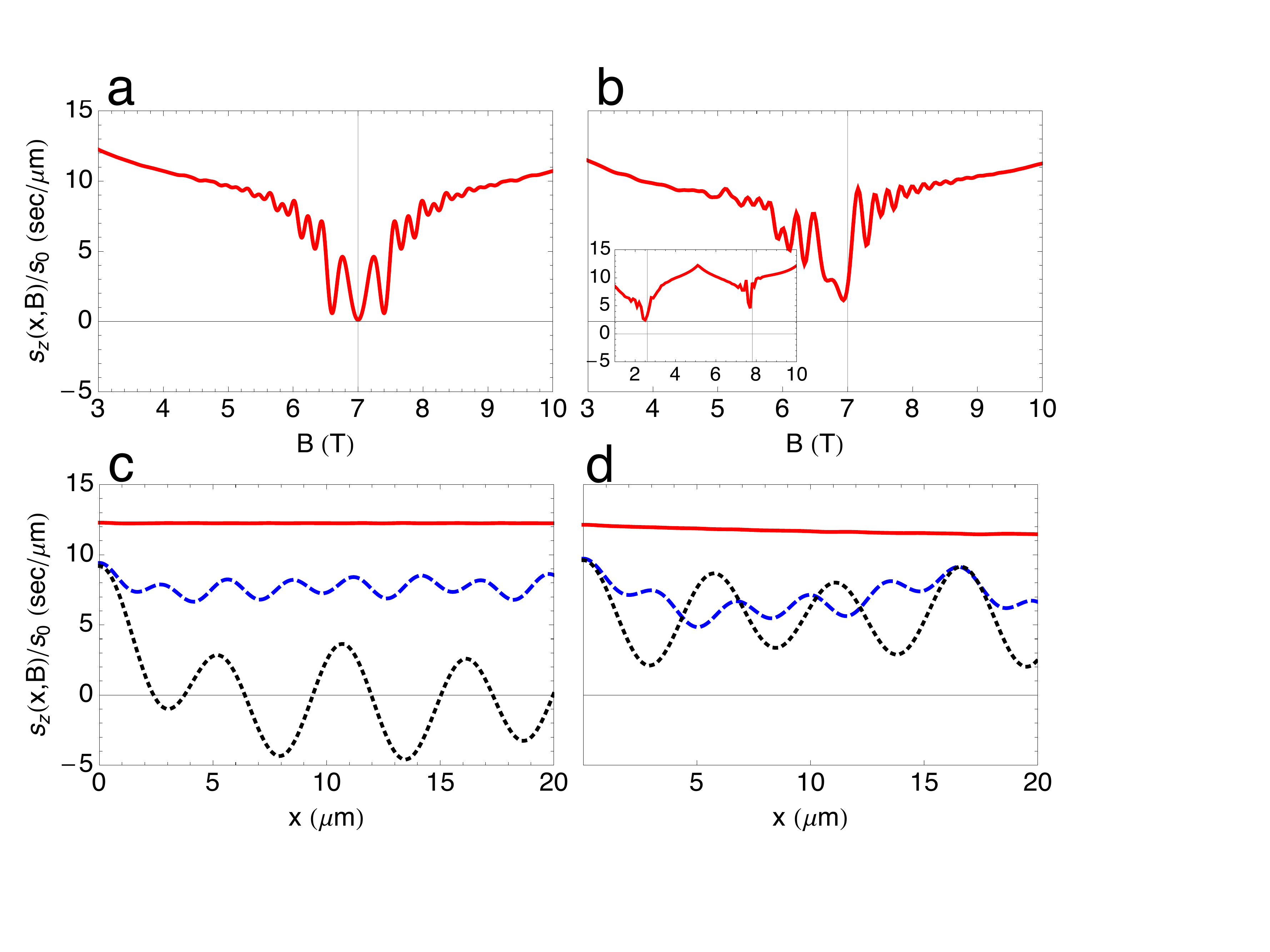}
 \caption{The stationary spin polarization $s_z$ as a function of the magnetic field $B$ for (a) parabolic and (b) square well lateral confinement.  In all computations $\alpha_- = 2 \times 10^{-13}$ eV-m and the width of the injection aperture is $0.5 \mu \mbox{m}$. The effective mass is $m = 0.067 m_e$ and the $g$ factor is $-0.44$.  The distance from the source, $x-x_s$, is $20 \mu \mbox{m}$ in panels a and b. For square well confinement the width of the well, $W$, in panels (b) and (d) is  $1.15 \mu \mbox{m}$  The inset in panel (b) shows $s$ vs $B$ when $W = 3 \mu \mbox{m}$ and  $ x = 6.7\mu \mbox{m}$ as in
 \cite{Frolov2009a}. Except for the inset, the number of propagating modes (including spin)  in all panels is 40. In the inset there are 105 propagating modes.
 Panels (c) and (d) show the dependence of the spin polarization on a distance from the source for the parabolic and square well lateral confinement respectively $B= 3$T (red,solid), $ B= 6 $T (blue,dashed), and  $B = 7$T (black, dotted). 
The parameters used for parabolic confinement are $\mu = 3.65$~meV, $\hbar\omega = 0.1785 $~meV, so that the magnetic field at the minimum is $7.0$~ T. For square well confinement  $\mu = 1.86$~meV, whereas in the inset it is $1.765$~meV.    Parameters have been chosen to mimic those of Ref.~\cite{Frolov2009a}.}
\label{fig:s_xB}  
 \end{figure}

\begin{figure}
\begin{center}
\includegraphics[width=1.\columnwidth]{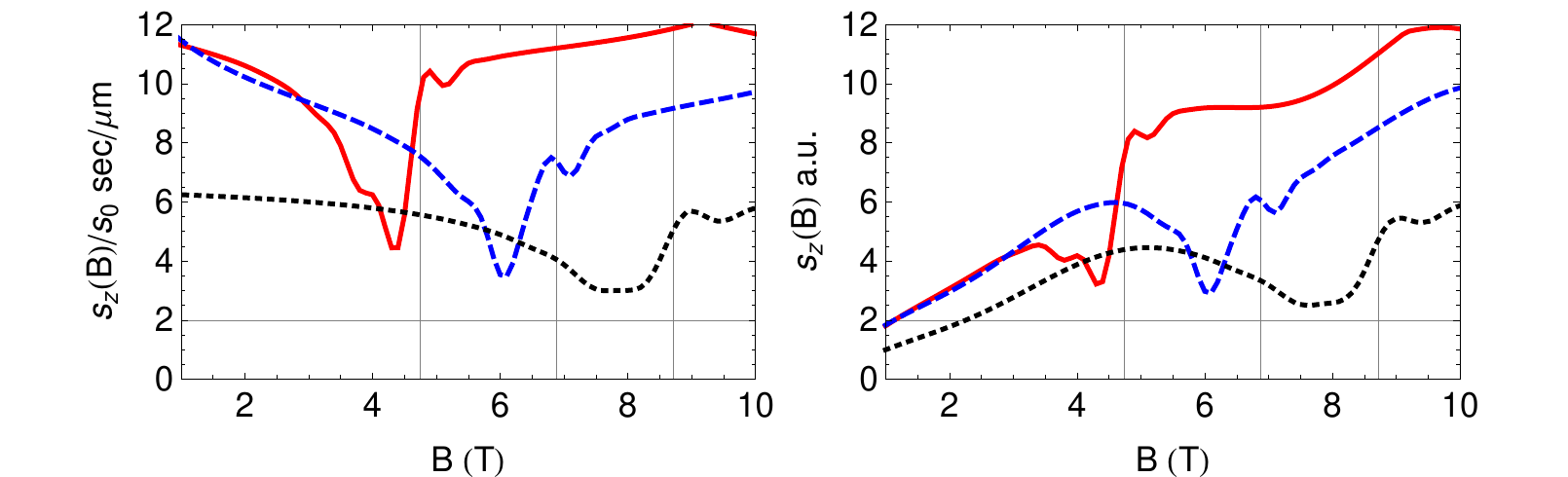}
\caption{ (Color online)
Spin polarization for 3 Fermi velocities:  solid(red) for $v_F = 0.7 \times 10^5$, dashed(blue) for $v_F = 1.0 \times 10^5 $, dotted for $v_F = 1.3 \times10^5$. Channel width is 1.2~$\mu$m. 
 }
\label{res2}
\end{center}
\end{figure}

The dependence of the off-diagonal matrix elements on the magnetic field is shown in 
Fig.~\ref{fig:matrix_elements}.
For parabolic confinement, modes at $ B=0 $ without SO splitting are equally spaced in energy; for a square well  the energy levels are spaced proportionally to $2 n +1$, where $n$ is the mode number. Ballistic spin resonance for a mode (spin) pair occurs when the Zeeman splitting is equal to the $B=0$ splitting.  Thus in the parabolic case all modes can become resonant, as indicated by the black diamonds in the left panel of Fig.~\ref{fig:matrix_elements}.  
In the square well case it is not possible for all modes to become resonant for a single value of $B$.  
Furthermore, in the square well, modes
that become resonant, but are located far away from the chemical potential, $\mu$, produce 
small effects on the spin polarization versus $B$ curves (see Fig.~\ref{fig:s_xB});  the dominant features of the spin resonance are determined by the spin-split mode pair with energies closest to $\mu$, i.e. the modes with the smallest $k_{n,\gamma}$.   These modes make the largest contribution to the spin density. 

The essence of the BSR is that the effective field becomes orthogonal to the injected spin polarization, i.e. $\cos \theta_n = 0.$  Any phase randomization (randomization of the $k_{n,\gamma}$) will then cause  decay of the polarization away from the injector. The behavior of the spin density without phase randomization is shown as a function of $B$ and as a function of $x$  in Fig.~\ref{fig:s_xB},  for parabolic and square well confinement. The shape of the resonance dip in panels (a) and (b) of Fig~\ref{fig:s_xB} depends on the nature of the confining potential.  It also depends on how close the modes are to cut-off; i.e on how small the $k_{n,\gamma}$ become.  In these figures we have plotted $s_z $ in a range of $B$ so that 
the uppermost modes are bounded away from cut-off. The impedance for injecting spins at the cut-off condition for a propagating mode must vanish, in analogy with total internal reflection in optics. The depth of the resonant dip depends on distance from the injection location $x-x_s$ as is shown in panels (c) and (d).
For both confinements,   $s_z$ vs. $x$  seems to be a sum of a small number of 
oscillating terms,  indicating  that only a few modes contribute to the resonant dips in panels (a) and (b). This has been verified (but is not shown here)   by simply calculating $s_z$ using only the two modes closest to cut-off. 
   
 The inset in Fig~\ref{fig:s_xB} shows $s$ vs $B$ having 2 minima. Parameters have been chosen to mimic Fig.~2b of Frolov~{\it et al.}~\cite{Frolov2009a}.  The first minimum comes from the condition 
 $ g |\mu_B| B = E_{m+1} - E_{m} = \Delta E_m $.  
 In the case of parabolic confinement only neighboring levels are coupled by the SO interaction.  In the case of square well confinement  SO interactions can couple  modes of the form (now the indices refer to states not coupled by SO) 
  $ | m ,s \rangle $ with modes    $ | m + (2n+1), -s \rangle $
 for any integer $n$.  For example with $ n=0 $, $ \Delta E_{m,1}= E_0 (2 m + 1) \approx   E_0 (2 m)$ while for  $n=1$,  $$ \Delta E_{m,3}= E_0 (6 m + 9) \approx   E_0 (6 m) = 3\Delta E_{m,1}.$$ Hence the corresponding values of $B_{resonant}$ are in the ratio of 3/1 as noted by Frolov~{\it et al.}~\cite{Frolov2009a}.

 Figure \ref{res2} illustrates the effect of changing the chemical potential $\mu$, or equivalently, the Fermi velocity $v_F = \sqrt{ 2  \mu/m}$.  For square well confinement,
 $E_n  = \hbar^2 \pi^2/ (2 m W^2) n^2 $ where $W$ is the width of the well. For large $n$ the resonance magnetic field is determined by 
 \begin{equation} 
 g \mu_B B_{res}\propto  2n \propto v_F.
 \end{equation} 
 The location of the minima in Fig.~\ref{res2} are roughly proportional to $v_F$ as expected from  the argument here.  In the right panel of Fig.~\ref{res2} we have multiplied each curve in the left panel by a function found by fitting the data for $B_x^{ext}$ found in Fig.~1c of Frolov et al.~\cite{Frolov2009a}. In their experiment, the rate of spin injection $s_0$ depends on the 
 magnetic field. This is proportional to the data  $s_x$ when the external field is in the x-direction, since in this case precession should not play a role in the observed spin ($x$) density.  If we assume that $s_0(B)$ is in fact independent of the direction of B, then we can use $s_0(B)$ to model spin-$z$  injection. The curves in the right panel of 
Fig.~\ref{res2} are qualitatively similar to the curves in Fig 2b of Frolov~{\it et al.}~\cite{Frolov2009a}.


\section{Effect of Disorder}
\label{sec:Effect}
The analysis in the previous section neglected the effect of disorder, which could disturb the spin-orbit entanglement upon which the resonant phenomena depends. To evaluate the role of the disorder in Ref.~\onlinecite{Frolov2009a} the typical mean free path 
$l $ and disorder-induced level broadening $\hbar / \tau $ should be compared to other length and energy scales  in the experiment. 
At a mobility of $4.5\times 10^6$cm$^2$/Vs,  $l \sim 20\mu$m and $\hbar/ \tau \sim 0.04$K.
The typical SO splitting is an order of magnitude larger, $\delta^{SO} \sim 0.4$K with the SO length $l_{SO} = 2\pi \hbar v_F / \delta^{SO} \sim 11.3\mu$m.
This definition of $l_{SO}$ is consistent with Eqs.~\eqref{2}, \eqref{Delta_n^SO} and \eqref{res:clean}.
The subband separation is $\Delta E \sim 1.5$K and thus the separate bands are well resolved ($\Delta E \gg \hbar / \tau$).
$\{ l_{SO}, W \} \ll l$, so the disorder broadening is the smallest energy scale in the problem and should not play a significant role in Ref.~\onlinecite{Frolov2009a}.
The remaining length scale, the source to drain separation, is comparable to $l$, and the diffusive nature of transport may lead to quantitative (but not qualitative) changes as we argue below.

The basic physical picture described by the two-subband model of Sec.~\ref{Introduction} also applies to the case of a  disordered channel.
The injected spin can be thought of as having  conserved and non-conserved (precessing) parts as in Fig.~\ref{precession}.
The conserved part diffuses along the channel rather than propagating ballistically at long distances.
If the source to drain separation $|x_s - x_d|\gg l$ the stationary 
polarization depends linearly on the distance $x$ for  $x_s < x < x_d$, see Fig.~\ref{setup}.
In our (effectively) one-dimensional model the diffusion equation in steady state gives $\partial_x^2 s_z = 0$,
which along with the boundary conditions at the source  and drain,
\begin{align}
 - D \partial_x ( x = x_s + 0^+) + D \partial_x ( x = x_s - 0^+) = s_0, \\
 - D \partial_x ( x = x_d + 0^+) + D \partial_x ( x = x_d - 0^+) =- s_0, 
\end{align} 
can be trivially solved to determine the spin polarization in a steady state.
Outside the region from injector to detector the polarization $s_z \propto \left|x_s - x_d \right|/  l $, which we support with a microscopic analysis in  Sec.~\ref{app:dis} and \ref{app:dis_weak}.

These arguments apply if the disorder conserves not just the spin, but also the pseudospin of a state. The condition for preserving the pseudospin is that the disorder potential doesn't mix the states $|n,\gamma\rangle$ defined in Eq.~\eqref{ball_eig} with opposite $\gamma$'s.
We show below that this condition is satisfied for a generic scattering potential.

The non-conserved part of an injected polarization oscillates on a scale $\lesssim l_{SO}$ away from the source (see Fig.~\ref{fig:s_xB} where the oscillation scale is about 5~$\mu$m).
Weak disorder with $\hbar/ \tau \ll \delta^{SO}$ is ineffective at experimentally relevant distances from the source, $|x-x_s|\lesssim l$.
Therefore the oscillations survive this weak disorder.
The conserved part of polarization can even be enhanced  by weak disorder, provided $|x_s -x_d| \gtrsim l$.
In Ref.~\onlinecite{Frolov2009a} $|x_s -x_d| \simeq l$ and this enhancement is not pronounced.
However, making the injector to detector separation larger may increase the magnitude of the disorder-driven enhancement of the conserved part of the polarization.

If either the detector is placed close to the node of spin oscillations shown in Fig.~\ref{fig:s_xB}(c,d), or the oscillations are smeared out due to the finite size of the injector and detector [as shown below in Fig.~\ref{fig:smooth}(c,d)] the measured signal is determined by the conserved part of the injected spin.
Thus beyond the distance $l_{SO}$ from the source, the $B$ dependence of a detected spin polarization is chiefly determined 
by the depolarization factor ($\cos^2 \theta$) of Eq.~\eqref{cos_n}.
Neglecting the variation of $\theta$ over bands,
\begin{align}\label{Lorentz}
s_z(x,B) \propto \cos^2 \theta \approx 1 - \frac{ (\delta^{SO})^2 }{ (E_z - \hbar \omega )^2 + (\delta^{SO})^2 } \, .
\end{align}
The spin polarization above drops to zero when $B - B_{res}$ vanishes, with the spin polarization following a Lorentzian curve with a
scale of variation of $\delta^{SO} \sim 0.4$K, which translates into a width of the resonance in magnetic field  $\sim 2.3$T.
The experimental data confirms this estimate.

We support these general statements with a microscopic analysis for  the case of strong disorder, $l \ll l^{SO}$, in Sec.~\ref{app:dis} and the case of weak disorder, $l \gg l^{SO}$, in Sec.~\ref{app:dis_weak}.  
In both cases the polarization follows Eq.~\eqref{Lorentz} at distances exceeding $ \min\{ l_{SO}, \sqrt{ l_{SO} l }\}$, at which point the non-conserved part of polarization drops.
For $l \gg l^{SO}$ the polarization decay is ballistic-like, and for $l \ll l^{SO}$
the polarization follows Hanle-like relaxation \cite{Johnson1985,Jedema2002,Frolov2009b}.


\subsection{Spin polarization in the presence of strong disorder, $W \ll l \ll l_{SO}$}
\label{app:dis}
In this section we derive expressions for spin correlation functions in the presence of the disorder with the mean free path larger than the channel width but smaller than the SO length.
We then use these correlation functions to obtain the result Eq.~\eqref{Lorentz}.
The quasi-classical approximation is justified provided
$k_F W , k_F l \gg 1 $.
The former condition is equivalent to having many propagating modes.
In addition we assume that the individual modes are well resolved, $W\ll l$.

Since the details of the disorder potential $V_{imp}(\vec{r})$ are not essential we assume it to be Gaussian with the simplest form of the correlation function,
\be\label{potential1}
\langle V_{imp}(\vec{r}) V_{imp}(\vec{r}') \rangle = V(z_1) V(z_2) \delta( x - x')\, .
\ee
Finally we simplify the analysis by neglecting the dispersion in angles $\theta_n$ of the effective magnetic field, i.e. we set $\theta_n \equiv \theta$.
This assumption is strictly satisfied at the BSR for a parabolic confinement potential, $\theta_n = \pi/2$.
It is also satisfied off the resonance when all $\theta_n \approx 0$.
In general a weak dispersion of angles $\theta_n$ is expected to cause D'yakonov-Perel' -like relaxation.

\begin{figure}[h]
\begin{center}
\includegraphics[width=1.0\columnwidth]{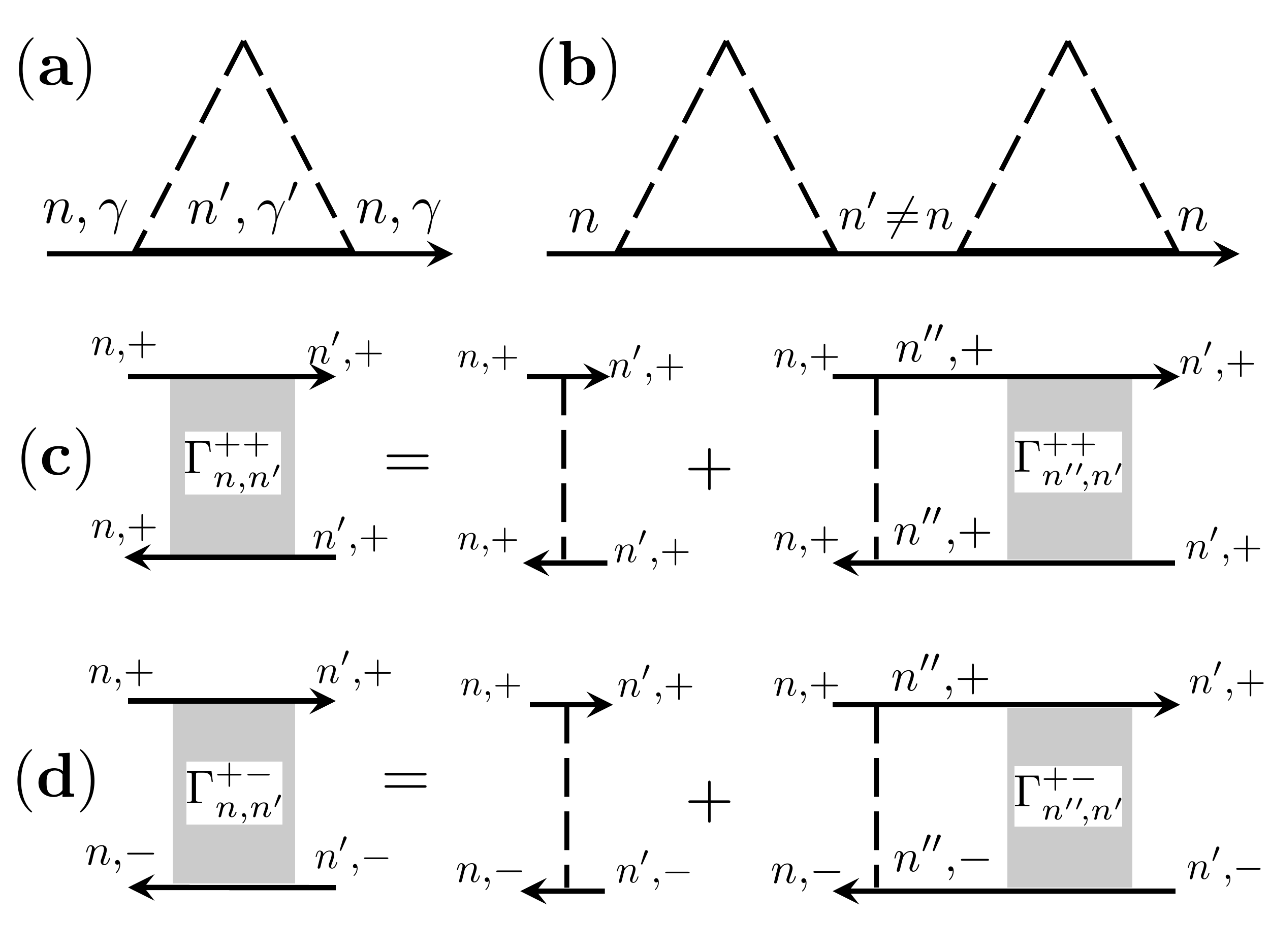}
\caption{ The diagrammatic representation of (a) the self consistent self energy, Eq.~\eqref{self}, (b) an example of a subleading in $(\Delta E \tau)^{-1}$ contribution.
Panel (c) [(d)] illustrates the Dyson equation, Eq.~\eqref{Gamma} satisfied by a scattering amplitude  $\Gamma^{++}_{n,n'}$ ($\Gamma^{+-}_{n,n'}$).
Solid arrowed lines denote the Green's functions.
The dashed line stand for disorder averaged scattering amplitudes.}
\label{fig:diag}
\end{center}
\end{figure}

We start with the consideration of the Green's function in the self-consistent Born approximation (SCBA), see Fig.~\ref{fig:diag}.
At each Fermi point we linearize the dispersion relation and introduce left and right moving species, labeled by the index $r = \pm 1$.
We are looking for the Green's function in the form
\begin{align}\label{disG1}
\hat{G}_{n,\gamma;r} (\epsilon,p) = 
| n,\gamma \rangle \langle n,\gamma |  G_{n,\gamma;r} (\epsilon,p),
\end{align}
where the ballistic eigenstates $|n, \gamma \rangle$ are introduced in Eq.~\eqref{ball_eig}.
\begin{align}\label{disG2}
G_{n,\gamma;r} (\epsilon,p) =
\frac{ 1 }{ \epsilon - r v_n p  - \gamma \Delta_n/2 + i \hbar / 2 \tau_n }\, ,
\end{align}
where $\Delta_n$ is the band splitting, Eq.~\eqref{Delta_n},  and $v_n$ is the Fermi velocity of the $n$th band.
The rate $\tau_n^{-1}$ is determined below within the SCBA using the correlation function  Eq.~\eqref{potential1}. 
The self-consistency condition is presented graphically in Fig.~\ref{fig:diag}(a).
In the case of well-resolved bands, $k_F l \gg 1$ or equivalently $\Delta E \tau/ \hbar \gg 1$, the Green's function Eq.~\eqref{disG2} is diagonal in the band index $n$.
Indeed, the self energy diagram presented in Fig.~\ref{fig:diag}(b) is smaller than the leading contribution shown in Fig.~\ref{fig:diag}(a) by a factor of $\Delta E \tau \gg 1$.
This justifies labeling of the Green's function by a single band index $n$.
We therefore retain only contributions to the self energy shown in Fig.~\ref{fig:diag}(a).
The self consistency condition represented graphically in Fig.~\ref{fig:diag}(a)  with the correlation function Eq.~\eqref{potential1} reads
\begin{align}\label{self}
\frac{ 1 }{ 2 \tau_n^{\gamma} } = 
\Im \int \frac{d p'}{ 2 \pi} \sum_{n',\gamma',r'} 
\left| V_{n,n'}^{\gamma \gamma'} \right|^2 
G_{n',\gamma';r'} (i\epsilon_m,p+p') \, ,
\end{align}
where the disorder matrix elements are 
\begin{align}\label{Vnn}
V_{n,n'}^{\gamma \gamma'} = 
\left\langle n, \gamma \left| V(z) \right| n' \gamma' \right\rangle \, .
\end{align}
With the explicit expressions for the ballistic eigenstates 
of the main text and using the fact that the potential does not flip the spin we write
\begin{align}\label{m_elements}
V_{n,n'}^{++} & = \sin \frac{ \theta_n}{2} \sin \frac{ \theta_{n'} }{2} V_{n+1,n'+1}
+ 
\cos \frac{ \theta_n}{2} \cos \frac{ \theta_{n'} }{2} V_{n,n'}
\notag \\
V_{n,n'}^{--} & = \cos \frac{ \theta_n}{2} \cos \frac{ \theta_{n'} }{2} V_{n+1,n'+1}
+
 \sin \frac{ \theta_n}{2} \sin \frac{ \theta_{n'} }{2} V_{n,n'}
 \notag \\
V_{n,n'}^{+-} & = - i \sin \frac{ \theta_n}{2} \cos \frac{ \theta_{n'} }{2} V_{n+1,n'+1}
+ 
i \cos \frac{ \theta_n}{2} \sin \frac{ \theta_{n'} }{2} V_{n,n'}
 \notag \\
V_{n,n'}^{-+} & =  i \cos \frac{ \theta_n}{2} \sin \frac{ \theta_{n'} }{2} V_{n+1,n'+1}
- 
i \sin \frac{ \theta_n}{2} \cos \frac{ \theta_{n'} }{2} V_{n,n'}\, ,
\end{align}
where
\begin{align}
V_{n,n'} = \int d z \phi_n(z) V(z) \phi_{n'}(z) \, .
\end{align}
Because of the large number of propagating modes, $V_{n+1,n'+1} \approx  V_{n,n'} $.
And since we assumed $\theta_n \approx \theta$, Eq.~\eqref{m_elements} simplifies to 
\begin{align}\label{m_elements1}
V_{n,n'}^{++} = V_{n,n'}^{--} =  V_{n,n'}\, ,
\quad
V_{n,n'}^{+-} = V_{n,n'}^{+-} = 0\, .
\end{align}
With Eq.~\eqref{m_elements1}, evaluation of Eq.~\eqref{self} is straightforward with the result,
\begin{align}\label{tau}
\frac{ \hbar }{ 2 \tau_n } = \sum_{n'} \frac{ V_{n,n'}^2 }{ \hbar v_{n'}}\, ,
\end{align}
where the sum runs over the propagating modes.
One can check that the self energy is diagonal in the band index.
This follows from the cancelation of $\gamma' = \pm 1$ contributions for the off diagonal part of the self-energy matrix.
Therefore the assumption of self energy being diagonal in both $n$ and $\gamma$ indices is shown to be self consistent.
This situation is very similar to the case of a constant Zeeman field, where the form of Eq.~\eqref{disG2} is known to hold.

We next consider collective density dynamics by analyzing the two-particle correlation functions.
Similar to the theory of disordered Fermi liquids\cite{Finkelstein1984}, the spin correlation function can be decomposed as a sum of a static and dynamic parts, 
$\chi = \chi^s + \chi^d$.
The static part $\chi^s$ contains contributions to the particle-hole ladder diagrams, Fig.~\ref{fig:diag}, in which the both upper and lower Green functions in Fig.~\ref{fig:diag}(c,d) are either both retarded or both advanced.
The $\chi^s$ is insensitive to disorder and other energy scales and reduces to a density of states at the Fermi energy,
$\chi_s(q,\omega) \approx \nu = 2 \sum_n (\hbar \pi v_n)^{-1}$.
The dynamic part $\chi^d$ consists of the contributions with upper(lower) Green functions of a ladder in Fig.~\ref{fig:diag}(c,d) being retarded(advanced). 

We now turn to the analysis of the related retarded-advanced (RA) scattering amplitude.
Because of  Eq.~\eqref{m_elements1} 
 two separate scattering amplitudes can be introduced, $\Gamma^{\pm\pm}(q,\omega)$ and   
$\Gamma^{\pm,\mp}(q,\omega)$, describing the diffusion of the conserved and precessing spin components respectively.
These amplitudes satisfy the corresponding Dyson equations presented graphically in Fig.~\ref{fig:diag}(c),(d)
\begin{align}\label{Gamma}
\Gamma^{\gamma\gamma'}_{n,n'} &= 
V^2_{n,n'} +  \sum_{n''} V^2_{n,n''} \Pi^{\gamma\gamma'}_{n''} \Gamma^{\gamma\gamma'}_{n'',n}\, ,
\end{align}
where the polarization operator 
\begin{align}\label{Pi1}
\Pi^{\gamma\gamma'}_{n}(q,\omega) = 
\sum_{r=\pm} \int \frac{ d p }{ 2 \pi } 
G^R_{n,\gamma;r} (\epsilon+\omega,p+q) G^A_{n,\gamma';r} (\epsilon,p)
\end{align}
is easily evaluated with Eq.~\eqref{disG2},
\begin{align}\label{Pi2}
\Pi^{\gamma\gamma'}_{n}(q,\omega) = &
\sum_{r=\pm}
\frac{ \tau_n}{v_n} \left[ 1 - i \omega \tau_n + i r v_n q \tau_n 
\right.
\notag \\
& + 
\left.
\frac{i}{2} (\gamma - \gamma') \Delta_n \tau_n \right]^{-1}\, . 
\end{align}
We address first the dynamics of the component parallel to the effective field by setting $\gamma = \gamma'$.
At long distances the spin density is dominated by  the spin diffusion mode away from the source.
To identify this mode we adopt the approach of Ref.~\cite{Wolfle1984}, and introduce the spectral decomposition of the scattering amplitude as follows,
\begin{align}\label{eig}
\sum_{n'} V_{n,n'}^2 \Pi^{\pm\pm}_{n'}(0,0)\varphi_{n'}^l = \lambda_l \varphi_n^l  \, .
\end{align}
Here the index $l = 0,1,\ldots$ labels eigenfunctions $\varphi_n^l $ normalized by the condition
\begin{align}\label{norm}
\sum_{n} \varphi_n^l \Pi^{\pm\pm}_{n}(0,0)\varphi_{n}^l = \delta_{l,l'}  \, .
\end{align}
Substituting the expansion $ V_{n,n'}^2 = \sum_l a^l_n \varphi_{n'}^l$ into Eq.~\eqref{eig} and using Eq.~\eqref{norm} we obtain  
\begin{align}\label{V}
V_{n,n'}^2 = \sum_l \lambda_l \varphi_{n}^l \varphi_{n'}^l \, .
\end{align}
The amplitude in Eq.~\eqref{Gamma} can be decomposed using the Eqs.~\eqref{eig}, \eqref{norm} and \eqref{V} as follows
\begin{align}\label{Gamma_spectral}
\Gamma^{\pm\pm}_{n,n'} = \sum_l \frac{ \lambda_l }{ 1 - \lambda_l }\varphi^l_n \varphi^l_{n'} \, .
\end{align}
The hydrodynamic mode is identified as the one having the eigenvalue $\lambda_{l=0} \approx 1$.
Following Ref.~\cite{Wolfle1984} this mode is written as
\begin{align}\label{0mode}
\varphi_n^{l = 0} = \frac{ Z }{ 2 \tau_n }\, , \quad Z = \sqrt{2} \left( \sum_n v_n^{-1} \tau_n^{-1} \right)^{-1/2}\, ,
\end{align}
where $Z$ is the normalization factor fixed by the condition \eqref{norm}. 
Indeed because it follows from Eq.~\eqref{Pi2} that $\Pi^{\pm\pm}(0,0) = 2 \tau_n / v_n$,
and because of the Eq.~\eqref{tau}, the definition  Eq.~\eqref{eig} gives 
\begin{align}\label{0modeA}
\sum_{n'} V_{n,n'}^2 \Pi^{\pm\pm}_{n'}(0,0)\varphi_{n'}^{l=0} =  \varphi_n^{l=0}  \, .
\end{align}
Equation \eqref{0modeA} shows that $\varphi^{l=0} $  corresponds to the hydrodynamic pole of the scattering amplitude Eq.~\eqref{Gamma_spectral} at $q=0$ and $\omega=0$.
This pole is a consequence of the conservation of the spin polarization parallel to the effective field.
To obtain the dynamics in long wavelength, i.e. hydrodynamic,  limit we analyze the small $q$ and small $\omega$ dependence of the $\lambda^{l=0}(q,\omega)$ eigenvalue.
At non-zero $q$ and $\omega$  the eigenvalue equation equation is
\begin{align}\label{0modeB}
\sum_{n'} V_{n,n'}^2 \Pi^{\pm\pm}_{n'}(q,\omega)\varphi_{n'}^{l=0} = \lambda^{l=0}(q,\omega) \varphi_n^{l=0}  \, .
\end{align}
In the hydrodynamic regime, $\omega \tau_n, v_n q \tau_n \ll 1$ the difference 
\begin{align}\label{diff}
\Pi^{\pm\pm}_{n}(q,\omega) - \Pi^{\pm\pm}_{n}(0,0) 
\approx
\frac{ 2 \tau_n}{ v_n } \left( i \omega \tau_n + v_n^2 q^2 \tau_n^2 \right)
\end{align}
can be considered as a small perturbation of an original eigenvalue problem at $q=0$, $\omega=0$ with the unperturbed solution Eq.~\eqref{0mode}.
The corresponding perturbation theory has been worked out in Ref.~\cite{Wolfle1984} and here we quote the result,
\begin{align}\label{q_omega}
\lambda^{l=0}(q,\omega) \approx 1 + i \omega \langle \tau^{-1} \rangle^{-1} - \langle D \rangle q^2 \langle \tau^{-1} \rangle^{-1}\, ,
\end{align}
where the averaging over modes is defined as a sum weighed by the density of states,
\begin{align}\label{average}
\langle A \rangle   = \frac{ \sum_n v_n^{-1} A_n }{ \sum_n v_n^{-1} }
\end{align}
and $D_n = v_n^2 \tau_n$.
In the quasi-classical regime the number of modes is large, and the sum in Eq.~\eqref{average} can be replaced by the integral,
\begin{align}
\label{average1}
\left\langle A(n)  \right\rangle_n \approx 
 \frac{ \int_{-k_F}^{k_F} d k A(k) (k_F - k^2)^{-1/2} }{ \int_{-k_F}^{k_F} d k (k_F - k^2)^{-1/2}  }\, ,
\end{align}
where $k_F$ is the bulk Fermi momentum.
Equation \eqref{average1} in turn corresponds to an angular integration over the Fermi surface with the energy independent density of states.
In the same limit the relaxation rate $\langle \tau \rangle^{-1}$ and the diffusion coefficient $\langle D \rangle $ approach their corresponding bulk values $1/ \tau$ and $k_F^2/ m^2 ( \tau/2)$ for a two-dimensional electron gas in the presence of $\delta$-correlated disorder.
In the case of the diffusion coefficient the numerical prefactor appears as a result of the identity 
$$ 
\frac{\int_{-1}^{1} d x x^2 (1 - x^2)^{-1/2} }{  \int_{-1}^{1} d x  (1 - x^2)^{-1/2} } = 1/2\, .
$$
Keeping only the hydrodynamic mode in the expansion \eqref{Gamma_spectral} as appropriate at large distances from the spin source we finally obtain
\begin{align}\label{Gamma_hydro}
\Gamma^{\pm\pm}_{n,n'} \approx  \frac{ \varphi^{l=0}_n \varphi^{l=0}_{n'} \langle \tau^{-1} \rangle }{ i \omega - \langle D \rangle q^2  },
\end{align}
which is natural for diffusion without spin precession and spin flip processes.


The diffusion of non-conserved component of the spin polarization is accompanied by oscillation.
This manifests itself in polarization operators at small $q$ and $\omega$,
\begin{align}\label{diff2}
\Pi^{\pm\mp}_{n}(q,\omega) - \Pi^{\pm\mp}_{n}(0,0) 
\approx
\frac{ 2 \tau_n}{ v_n } \left( i \omega \tau_n + v_n^2 q^2 \tau_n^2  \mp i \Delta_n \tau_n \right)
\end{align}
which differs from Eq.~\eqref{diff} by the presence of a spin splitting contribution.
Similarly to \eqref{Gamma_hydro} we obtain 
\begin{align}\label{Gamma_hydro2}
\Gamma^{\pm\mp}_{n,n'} \approx  \frac{ \varphi^{l=0}_n \varphi^{l=0}_{n'}  \langle \tau^{-1} \rangle }{ i \omega  - \langle D \rangle q^2  \mp i \langle \Delta \rangle }\, .
\end{align}
The most singular contributions to a dynamical spin susceptibility 
\begin{align}\label{sing}
\chi^d(q,\omega) \approx & \frac{ i\omega }{ 2 \pi } \sum_{\gamma=\pm} \sum_{n,n'}
\left[
\cos^2 \theta \,   
 \Pi_{n}^{\gamma,\gamma} \Gamma^{\gamma,\gamma}_{n,n'}  \Pi_{n'}^{\gamma,\gamma}
\right. 
 \notag \\
& \left. + 
\sin^2 \theta \,
\Pi_{n}^{\gamma,-\gamma} \Gamma^{\gamma,-\gamma}_{n,n'}  \Pi_{n'}^{\gamma,-\gamma} 
\right]\, ,
\end{align}
where the $\cos\theta$ and $\sin\theta$ are the matrix elements of the spin operator in the $|n,\pm\rangle$ basis introduced in the discussion of the ballistic case.
In  equation \eqref{sing} the polarization operators can be evaluated at $q=0$ and $\omega=0$, so that from Eq.~\eqref{Pi2}, $\Pi^{\gamma\gamma'}_n \approx 2 \tau_n / v_n$.
Then substitution of \eqref{0mode}, \eqref{Gamma_hydro} and \eqref{Gamma_hydro2} into Eq.~\eqref{sing} gives
\begin{align}\label{sing1}
\chi^d(q,\omega)& \approx \nu  \cos^2 \theta  \frac{ \omega }{ \omega  + i \langle D \rangle q^2  }
\\
+&
 \frac{ \nu }{ 2 }  \sin^2 \theta
 \left(
 \frac{  \omega }{ \omega - \langle \Delta \rangle + i \langle D \rangle q^2  }
+
  \frac{  \omega  }{  \omega  + \langle \Delta \rangle + i  \langle D \rangle q^2  } \right)\, . \notag
\end{align}
The first term in this equation describes diffusion of the spin component parallel to the effective field.
The second term describes the diffusion of oscillating spin polarization (coming from precessing pseudospins).
The stationary spin polarization obtained by a substitution of Eq.~\eqref{sing1} in Eq.~\eqref{Main3} reads
\begin{align}\label{illustrate}
s_z(x) & \approx 
s_0 \cos^2 \theta \left( \left|x_s - x_d \right|/ 2 \langle D \rangle \right)
\\
- &
\frac{ s_0 l_H }{2^{3/2} \langle D \rangle} \sin^2 \theta  
e^{- |x-x_s|/l_H}
 \cos\left( \frac{ \pi}{ 4 } + \frac{  |x-x_s|}{ l_H } \right) \, , \notag
\end{align}
where the Hanle length $l_H = \sqrt{2 \hbar \langle D \rangle/ \langle \Delta \rangle } = \sqrt{2 l_{SO} l} \gg l$.
For simplicity we assumed in Eq.~\eqref{illustrate},
$|x_s - x_d| \gg l_H \gg l$. 
At $x- x_s \gg l_H$ the second term in \eqref{illustrate} is negligible and the result \eqref{Lorentz} is obtained using the definitions \eqref{cos_n} and \eqref{sin_n} and replacing $\delta_n^{SO} \approx \delta^{SO}$.

In the above derivation the dispersion in $\theta_n$ angles has been neglected.
We expect the variation of $\theta_n$ to induce randomness in the effective  field and therefore cause spin relaxation akin to a D'yakonov-Perel' mechanism of spin relaxation.
Nevertheless the corresponding relaxation rate is expected to be suppressed relative to its bulk value because such band to band variations of $\theta_n$ are rather small.
In parabolic confining potential at the resonance $\theta_n = \pi /2$ for all $n$.
Likewise, off the resonance $\theta_n \approx 0$.
This spin relaxation is left for future studies.
\subsection{Spin polarization in the presence of weak disorder, $W \ll l_{SO} \ll l$}
\label{app:dis_weak}

Weak disorder has little effect on the non-conserved (precessing) part of spin polarization, whereas 
the conserved part still diffuses along the channel for distances exceeding the mean free path.
Therefore  the second term of Eq.~\eqref{res:clean}, which is valid in the ballistic case, applies for the non-conserved component and the first term of Eq.~\eqref{illustrate} is appropriate for the conserved part, which is still determined by a simple diffusion equation. 
For distances much more than $l_{SO}$ away from the source the signal is dominated by the conserved part and the BSR effect is qualitatively described by Eq.~\eqref{Lorentz}.

To justify this we begin with  Eq.~\eqref{Pi1}, which remains valid.
The components of the polarization operator $\Pi^{\gamma\gamma'}_{n}(q,\omega)$
with $\gamma \neq \gamma'$ contain the  factor $\Delta_n \tau_n$ in the denominator, where $\Delta_n \tau_n\gg 1$, [Eq.~\eqref{Pi2}]:
\begin{align}\label{estimate1}
\Pi^{\pm \mp}_{n}(q,\omega) = &
\sum_{r=\pm}
\frac{ \tau_n}{v_n} \left[ 1 - i \omega \tau_n + i r v_n q \tau_n 
\pm i  \Delta_n \tau_n \right]^{-1}\, . 
\end{align}
In the interval $x \gg l_{SO}$ we can approximate Eq.~\eqref{estimate1} as
\begin{align}\label{estimate2}
\Pi^{\pm \mp}_{n}(q,\omega) \simeq
 \pm 2 i 
\frac{ \tau_n}{v_n} 
\left( \Delta_n \tau_n \right)^{-1}\, . 
\end{align}
From Eq.~\eqref{tau} we can estimate the scattering vertex  
$V^2 \simeq \hbar v/{ \tau}$,
where the subband indices are omitted for clarity.
This yields an estimate of the product
\begin{align}\label{estimate4}
V^2 \Pi^{\pm \mp} \simeq \Delta \tau \ll 1\, .
\end{align}
It follows from Eq.~\eqref{estimate4} that the components $\Pi^{\pm \mp}$ are approximately their ballistic counterparts, as the Dyson series [Fig.~\ref{fig:diag}(d)] produces corrections to the ballistic result which are small by a factor $\sim(\Delta \tau/ \hbar)^{-1} = l_{SO} / l$.

The components  $\Pi^{\pm \pm}$, Eq.~\eqref{Pi2} with $\gamma = \gamma' = \pm$, do not contain the parameter $\Delta \tau$, so
\begin{align}\label{estimate5}
\Pi^{\pm \pm}_{n}(q,\omega) = 
\sum_{r=\pm}
\frac{ \tau_n}{v_n} \left[ 1 - i \omega \tau_n + i r v_n q \tau_n \right]^{-1}\, ,
\end{align}
which describes the diffusion of the conserved part regardless of the ratio $l/l_{SO}$.
Combining these observations we obtain for the steady state spin polarization
\begin{align}\label{res:clean_1}
s_z & (x,B) =  
s_0 \cos^2 \theta \frac{\left|x_s - x_d \right|}{ 2 \langle D \rangle }
\notag \\
&+ \frac{s_0}{2 \sum_{n} v_n^{-1}} \sum_{n} 
 v_n^{-2 } \sin^2 \theta_n  \cos \left[ \frac{2 \pi}{ l_n } (x - x_s) \right] \, .
\end{align}

\section{Discussion}
\label{sec:Discussion}
We have studied theoretically the phenomenon of BSR in quasi-one-dimensional channels.
Our theory reliably reproduces the drop in the non-local voltage induced by non-equilibrium spin polarization in both ballistic and diffusive regimes.
We argue that the BSR occurs due to  SO-induced depolarization.
When the Zeeman splitting is comparable to the energy scale of transverse quantization, the bands become doubly degenerate, as in  Fig.~\ref{2subband}.
By hybridizing these pairs of bands SO coupling causes the precession of the injected spin polarization as in Fig.~\ref{precession}.
In the ballistic regime the decay of the injected spin polarization is due to the variation in spin precession angle in various modes in the channel as well as the finite source size.
In the diffusive regime the dynamics of injected spins is a Hanle-like relaxation. 

In this paper we model the spin injection as a continuous influx of a quasi-equilibrium polarization, as in Ref.~\cite{Forster1975}.
While such an approximation may apply to a spin injection from a ferromagnetic contacts to a bulk \cite{Crooker2005b} its application to the injection via QPC has yet to be studied.
Experimentally the voltage is fixed, not the spin injection rate.
A more rigorous way of describing the injection would be to use a Landauer approach.
In a multi-terminal geometry this  requires solution of a quantum mechanical scattering problem for electrons injected via the QPC.

\begin{figure}
 \includegraphics[width=1.2\columnwidth]{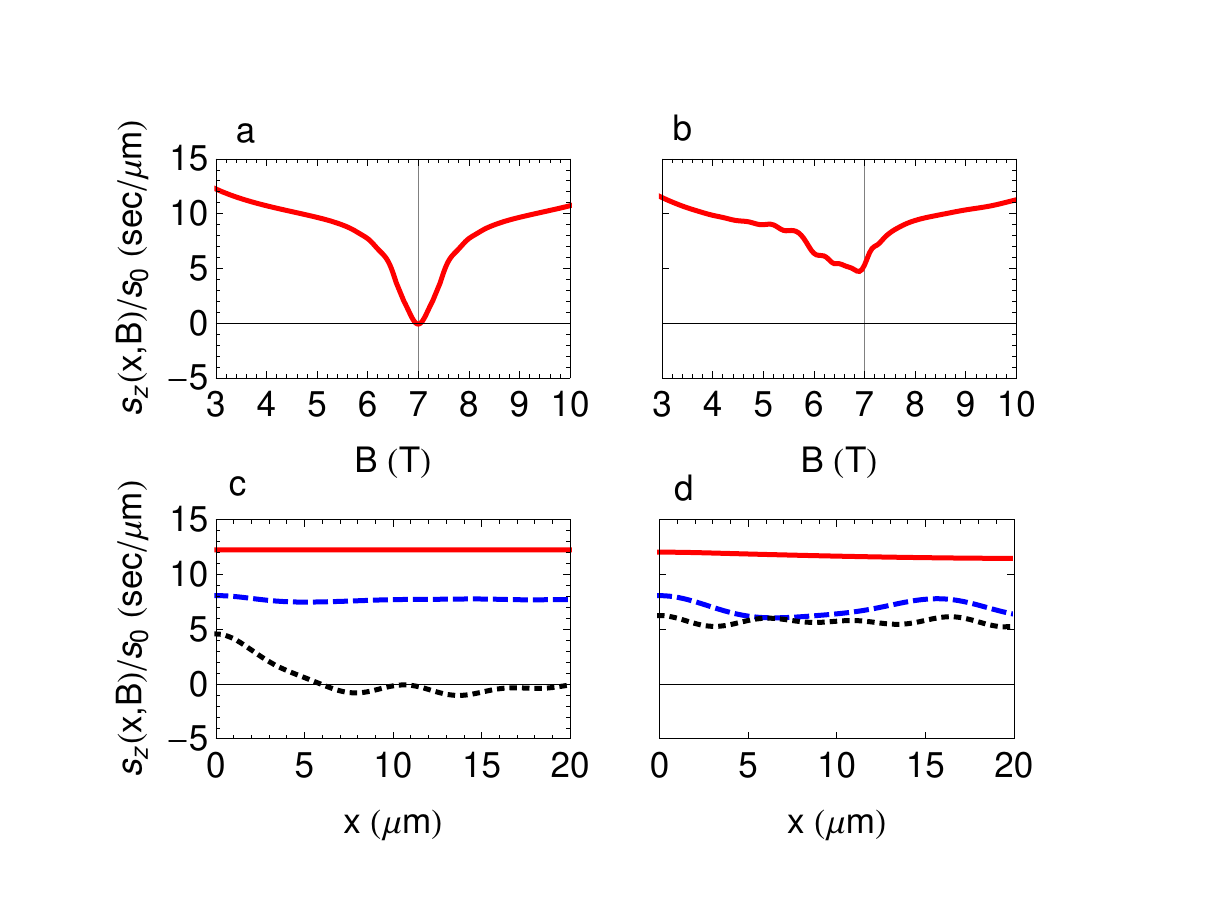}
 \caption{The same as Fig.~\ref{fig:s_xB} but with the effective spin injector size $2 \mu$m four times larger than $0.5\mu$m used in Fig.~\ref{fig:s_xB}.
In this case the effective injector size is comparable to the typical channel width $W = 1.2\mu$m.
The dip in spin polarization is more pronounced for the parabolic lateral confinement (a) as the resonance condition Eq.~\eqref{Resonant} can be satisfied for all bands simultaneously which is not the case for the infinite square well confinement (b).
This is due to the washing out of an oscillatory precessing polarization component with distance. 
The suppression of oscillations occurs for larger injector sizes and is present for both the parabolic (c) and square well (d) confinements.
}
\label{fig:smooth}  
 \end{figure}

Although we have captured the gross features of BSR, a more accurate modeling of a spin injection is necessary for a detailed description of spin transport in channels contacted by a QPC.
For example, when a band is just touching the Fermi level, 
the velocity of the carriers along the channel  is nearly zero, so that if the injection rate is kept constant a divergent spin accumulation would result.
The injection rate should instead become zero due to perfect reflection at the QPC, which eliminates the divergence.
We also note that the effective size of an injection or detection region depends on the microscopic details of a QCP.
Within our theory therefore it should be regarded as a phenomenological parameter.
In Fig.~\ref{fig:s_xB} the injector size was taken to be $0.5 \mu$m.
If one takes $2 \mu$m instead the spin polarization is noticeably smoothed out, as seen in Fig.~\ref{fig:smooth}.

Oscillations in non-local voltage due to ballistic spin precession have been reported in Ref.~\cite{Koo2009}.
The oscillations seen in our Fig.~\ref{fig:s_xB} reflect this observation.
Injection via the QPC differs from  injection from the long contact bar used in \cite{Koo2009}  and better resembles Fig.~\ref{fig:smooth}, with an effectively larger  injector and detector region.
A more detailed study of the spin injection across a QCP will be helpful to develop further understanding, and will be addressed in forthcoming studies.

The BSR phenomenon is an ideal setting to study  
the role of electron-electron interaction in spin transport and  spin relaxation in confined geometries, which was the primary motivation for this work. 
Of particular interest is an interaction induced renormalization of the resonant field which will be studied in the future based on the theory  above. Spin relaxation and coherence of the precessing pseudospins in the channel are likely to yield additional surprises.

\begin{acknowledgments}
We thank J.A. Folk and A.M. Finkel'stein for useful discussions. 
M.K. acknowledges the support of the University of Iowa. M.E.F. was supported in part by C-SPIN, one of six centers of STARnet, a Semiconductor Research Corporation program, sponsored by MARCO and DARPA.
\end{acknowledgments}

\begin{appendix}*
\section{Stationary spin polarization}
\label{sec:app1}

In this appendix we derive the general relation \eqref{Main3} assuming linear response, and illustrate it for simple diffusion.
For clarity we consider a steady influx of a specified particle density;
the number of particles imparted into the system per unit time and per unit volume is the flux $F_{\rho}(\vec{x})$.
In order to relate the steady state density to the retarded density correlation function $\chi_{\rho}(\vec{x},t)$ we rely on the following auxiliary result proven in Ref.~\cite{Forster1975}.
If an excess density $\rho(\vec{x})$ is prepared at an initial time instant $t_i$, the subsequent time evolution of the density at a given time $t$ and location $\vec{x}$, $\rho(\vec{x},t)$, is determined by the  correlation function $\chi_{\rho}(\vec{x},t)$ according to
\begin{widetext}
\begin{align}\label{lemma}
\rho(\vec{x}, t >t_i) = -i\int \frac{ d \omega }{ 2 \pi } 
\int d \vec{x}' \int d \vec{x}''
e^{-i \omega(t - t_i) }
\frac{\left[ \chi_{\rho}(\vec{x} - \vec{x}',\omega)  - \chi_{\rho,st}(\vec{x} - \vec{x}') \right]}{  \omega - i 0^+ }
\chi^{-1}_{\rho,st}(\vec{x}' - \vec{x}'') \rho(\vec{x}'',t_i)   \, ,
\end{align}
where the static correlation function $\chi_{\rho,st}(\vec{x}' - \vec{x}'') = \chi_{\rho}(\vec{x}' - \vec{x}'',\omega = 0)$.
During each infinitesimal time interval $[t_i, t_i+d t_i]$ the density 
$F_{\rho}(\vec{x}) d t_i$ is imparted into the system.
This density is evolving in time according to \eqref{lemma}.
In the linear response regime the separate contributions of each of the time intervals add up to yield 
\begin{align}\label{lemma1}
\rho(\vec{x}, t) =-i  \int_{-\infty}^t d t_i \int \frac{ d \omega }{ 2 \pi } 
\int d \vec{x}' \int d \vec{x}''
e^{-i \omega(t - t_i) }
\frac{\left[ \chi_{\rho}(\vec{x} - \vec{x}',\omega)  - \chi_{\rho,st}(\vec{x} - \vec{x}') \right] }{ \omega - i 0^+ }
\chi^{-1}_{\rho,st}(\vec{x}' - \vec{x}'') F_{\rho}(\vec{x}'')   \, .
\end{align}
For a flux that is turned on adiabatically,  
$F_{\rho}(\vec{x}'')$ is replaced in \eqref{lemma1} by $F_{\rho}(\vec{x}'',t_i) = F_{\rho}(\vec{x}'') \exp(\delta^+ t_i )$. The steady-state result is then obtained by taking the limit $\delta\rightarrow 0$ at the end of the calculation.
With this regularization the integration over the injection time $t_i$ is straightforward and yields the following expression:
\begin{align}\label{lemma2}
\rho(\vec{x}) = -\int \frac{ d \omega }{ 2 \pi } 
\int d \vec{x}' \int d \vec{x}''
\frac{\left[ \chi_{\rho}(\vec{x} - \vec{x}',\omega)  - \chi_{\rho,st}(\vec{x} - \vec{x}') \right] }{(  \omega - i 0^+)(  \omega - i \delta^+) }
\chi^{-1}_{\rho,st}(\vec{x}' - \vec{x}'') F_{\rho}(\vec{x}'')   \, .
\end{align}
The integration over the frequency $\omega$ can be done by closing the contour of integration in the upper half of a complex $\omega$ plane due to the fast decay of the integrand at large $\omega$.
It is essential that the retarded correlation function $\chi_{\rho}(\vec{x} - \vec{x}',\omega)$ is analytic in the area enclosed by the above contour.
Using the residue theorem and introducing spatial Fourier harmonics we rewrite \eqref{lemma2} as 
\begin{align}\label{lemma3}
\rho(\vec{x}) =
\left.
- i \int \frac{ d^d q }{ (2 \pi)^d }e^{ i \vec{q} \vec{x} } 
\left[ \partial_{\omega} \chi_{\rho}(\vec{q},\omega) 
\chi^{-1}_{\rho,st}(\vec{q}) \right] \right|_{\omega \rightarrow i \delta} F_{\rho}(\vec{q})   \, ,
\end{align}
where $d$ is the  dimension of space.
We note that the order of the limits taken with the regularizing infinitesimals $0^+$ and $\delta^+$ in \eqref{lemma3} is not important.

The arguments leading to Eq.~\eqref{lemma3} can be repeated for  spin injection.
The steady state spin polarization density $\vec{s}(\vec{x})$ is determined by the spin polarization rate $\vec{F}$ following a similar approach to Eq.~\eqref{lemma3}, yielding
\begin{align}\label{lemma4}
\vec{s}(\vec{x}) =
\left.
- i \int \frac{ d^d q }{ (2 \pi)^d }e^{ i \vec{q} \vec{x} } 
\left[ \partial_{\omega} \bar{\vec{\chi}}(\vec{q},\omega)
\bar{\vec{\chi}}^{-1}(\vec{q}) \right] \right|_{\omega \rightarrow i \delta} \vec{F}_{\vec{q}}   \, ,
\end{align}
where the $\bar{\vec{\chi}}(\vec{q},\omega)$ is the dynamical spin polarization tensor and  matrix multiplication is assumed.

To realize the specific case treated in this paper we assume  the injection rate $\vec{F}(\vec{x})$ to be constant across the channel, and introduce a spin density which has been integrated over the channel cross section,
\be\label{spin_op_vec}
\hat{\vec{s}}(x) = \frac{1}{2} \int_{-\infty}^{+\infty} d z 
\left[ \psi^{\dag}_{s_1} (x,z) \vec{\sigma}_{s_1,s_2}  \psi_{s_2}(x,z) \right]\, ,
\ee
where $\vec{\sigma}$ are the Pauli matrices and $s_{1,2}$ are dummy spin indices.
Integrating Eq.~\eqref{lemma4} over the cross-section of a quasi-one-dimensional channel and using the definition Eq.~\eqref{spin_op_vec} we obtain for the steady state value of the integrated spin density, 
\begin{align}\label{lemma5}
\vec{s}(x) =
\left.
- i \int \frac{ d q }{2 \pi }e^{ i q x } 
\left[ \partial_{\omega} \vec{\chi}(q,\omega)
\vec{\chi}^{-1}(q) \right] \right|_{\omega \rightarrow i \delta} \vec{F}_{q}   \, ,
\end{align}
where the integrated spin polarization tensor
\be\label{retcorr_app}
\chi^{ij}_{q,\omega} =  
- i \int_0^{\infty} d t \int  d x
e^{ - i q x + i \omega t }
\langle [\hat{s}_{i}(x,t),\hat{s}_{j}(0,0)] \rangle.
\ee

We now consider injection of a $z$-component of polarization, which is the experimental geometry, and set $\vec{F}(\vec{x}) = \hat{z} F(x)$.
The resulting $z$-component of the integrated spin polarization  from Eqs.~\eqref{lemma5} and \eqref{retcorr_app} is 
\begin{align}\label{lemma6}
s_z(x) =
\left.
- i \int \frac{ d q }{2 \pi }e^{ i q x } \sum_{j =x,y,z}
\left\{ \partial_{\omega} \chi^{zj}(\vec{q},\omega)
\left[\chi^{-1}(\vec{q})\right]^{jz} \right\} \right|_{\omega \rightarrow i \delta} F_{q}   \, .
\end{align}
For the geometry considered in the main text only $j=z$ provides non-zero contributions to Eq.~\eqref{lemma6}, for
although $\hat{s}_z(x)$ has non-zero matrix elements (only within the two dimensional eigenstate subspaces
$\alpha | n+1,\uparrow \rangle + \beta  | n, \downarrow \rangle $, $\beta^* | n+1,\uparrow \rangle - \alpha^*  | n, \downarrow \rangle $)
all four matrix elements in the same subspace of $\hat{s}_x(x)$ and $\hat{s}_y(x)$ are zero.
Therefore \eqref{lemma6} reduces to
\begin{align}\label{lemma7}
s_z(x) =
\left.
- i \int \frac{ d q }{2 \pi }e^{ i q x } 
 \partial_{\omega} \chi^{zz}(q,\omega)
\left[\chi^{-1}(q)\right]^{zz}\right|_{\omega \rightarrow i \delta} F_{q}   \, .
\end{align}
Finally by writing $\chi^{zz}(q,\omega) = \chi(q,\omega)$ and $\chi^{zz}(q) = \chi(q)$ Eq.~\eqref{Main3} is obtained.


To illustrate the result \eqref{lemma3} we consider  one-dimensional diffusion, which has a density correlation function 
\begin{align}\label{diff_rho}
\chi_{\rho}(q,\omega) = \nu \frac{ D q^2 }{ D q^2 - i \omega}\, ,
\end{align}
where $\nu$ is the density of states and $D$ is the diffusion coefficient.
Substitution of \eqref{diff_rho} in \eqref{lemma3} yields 
\begin{align}
\rho(x) =
\left.
\int \frac{ d q }{ 2 \pi }\frac{e^{ i \vec{q} \vec{x} }  F_{\rho}(q)  }{ D q^2 - i \omega }
 \right|_{\omega \rightarrow i \delta}  \, .
\end{align}
This equation can be written also in real space,
\begin{align}\label{diff1}
\rho(x) = \int_0^{\infty} d t \int_{-\infty}^{\infty} d x' D_c(x-x',t) F_{\rho}(x')\, ,
\end{align}
where the diffusion kernel,
\begin{align}\label{Dc}
D_c(x,t) = \frac{ \exp(- x^2 / 4 D t )}{ \sqrt{ 4 \pi D t } }\, .
\end{align}
Equations \eqref{diff1} and \eqref{Dc} can be easily generalized to include spin precession, drift and relaxation leading to, {\it e.g.}, Eq.~(1) of Ref.~\cite{Crooker2005b}.
\end{widetext}
\end{appendix}
\bibliographystyle{apsrevmod}

\end{document}